\documentclass[doublecol]{epl2} 

\usepackage{epsfig}
\usepackage{amsmath}
\usepackage{amssymb}

\usepackage[pdftitle={Zipping mechanism for force-generation by growing filament bundles},
            pdfauthor={Torsten Kuehne,Reinhard Lipowsky,  Jan Kierfeld},
        pdfkeywords={Zipping mechanism for force-generation by growing filament bundles}
            ]{hyperref}

\title{Zipping mechanism for force-generation by growing filament bundles}

\author{Torsten K\"uhne\inst{1} , Reinhard Lipowsky\inst{1}  
\and Jan Kierfeld\inst{1,2}}
\shortauthor{T. K\"uhne  \etal}

\institute{
 \inst{1} Max Planck Institute of Colloids and Interfaces, Science
  Park Golm, 14424 Potsdam, Germany, EU\\
  \inst{2} Physics Department,
  TU Dortmund University, 44221 Dortmund, Germany, EU
}


\abstract{
We investigate the force generation by polymerizing 
bundles of filaments, which form because of 
short-range attractive filament interactions. 
We show that bundles can generate forces by a 
zipping mechanism, which is not limited by buckling 
and operates in the fully buckled state.
The critical zipping force, i.e. the maximal force 
that a bundle can generate, is given by the 
adhesive energy  gained  during bundle formation.
For opposing forces larger than the critical zipping 
force, bundles undergo a  force-induced unbinding 
transition. For larger bundles, the critical zipping 
force depends on the initial configuration  of the bundles.
Our results are corroborated by Monte Carlo simulations. 
 }

\begin{document}

\maketitle

\section{Introduction}

Filamentous polymers play an important role in biological and chemical
physics. Both cytoskeletal filaments such as filamentous 
actin and microtubules and
chemically synthesized polymers such as dendronized polymers have diameters in
the range from 2 to 25 nanometers which leads to a considerable bending
rigidity, i.e.\ the persistence length is comparable or larger than the
polymer's contour length. 
The most important building blocks of the
cytoskeleton are actin  filaments with 
 a persistence length of $L_{p} \sim 15 \mu m$ 
and  microtubules  with a much larger 
 persistence length $L_{p} \sim 5 mm$.
Such semiflexible polymers 
are governed by several competing 
energy scales  in the system:
the bending energy and the thermal energy of the filaments, the interaction
energy between the filaments, and biochemical forces. 
In biological
systems, such biochemical
 forces are generated by  the activity of molecular motors proteins
 or the polymerization
dynamics of cytoskeletal filaments \cite{bray}.

Force generation by polymerizing cytoskeletal filaments is essential for 
various  cellular processes, such as motility \cite{bray}
 or the formation of cell protrusions including 
filopodia, lamellipodia, or acrosomal extensions 
\cite{theriot00,mogilner_rev}, where filaments push against a planar 
obstacle. 
Single 
polymerizing filaments can generate forces in the piconewton range, 
which  arise
from the gain in chemical bonding energy upon monomer 
attachment \cite{peskin93}. 
This process also involves shape fluctuations of the filament 
\cite{mogilner96}, which exerts entropic forces on the planar 
obstacle \cite{gholami06}.
Polymerizing filaments 
buckle at some critical length under the action of their own polymerization
force \cite{dogterom97}, which limits force generation by single filaments.

Filament bundles support cell protrusions and serve as stress
fibres \cite{bartles00,faix06}. 
Filament bundles have a higher bending rigidity 
and are, thus, 
more stable against buckling if a 
compressive load is applied \cite{bathe08}.
The formation of filament bundles is governed by the competition of 
thermal fluctuations and 
attractive interactions, which
can arise from crosslinking proteins or unspecific interactions.
Crosslinker-mediated interactions allow a reversible
formation of actin bundles, which can be regulated by the concentration of
crosslinkers in solution \cite{kkl05}.

\begin{figure}
  \begin{center}
  \epsfig{file=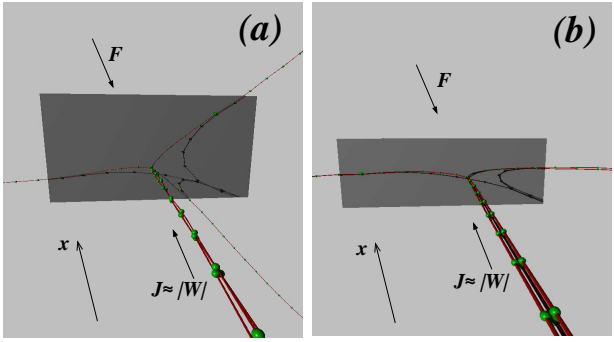,width=0.45\textwidth}
  \caption{
   Snapshots of MC simulations for $N=3$ filaments 
    close to the transition 
   between zipping and force-induced unbinding 
    for 
  two different initial conditions 
    (a)  $b=123$ and (b) $b=[12]3$.
}
  \label{figsnapshot}
  \end{center}
\end{figure}

Cellular force generating structures 
are typically made of  polymerizing bundles rather than single filaments. 
One reason is the enhanced stability of crosslinked 
stiffer bundles against buckling.
Moreover, ensembles of $N$ filaments  
could  share a compressive load  
force suggesting that the maximally generated force increases by a factor 
of $N$, similar to protofilaments in 
a microtubule \cite{dogterom00}. 
In addition,  crosslinking  within  
filament bundles can allow the bundle to generate 
higher forces  by exploiting the additional interaction energy 
\cite{mahadevan00,MO03}. 
As a result,  the mechanism of force generation by polymerizing bundles is 
difficult to understand because it  involves
several types of forces: 
chemical polymerization forces from monomer bonding, 
entropic forces from shape fluctuations, and interaction forces.
Moreover, a critical buckling force limits the mechanical stability
of filaments.
In this Letter, we show that there exists one  possible mechanism 
of force generation by filament bundles,
 the so-called {\em zipping mechanism},
which is completely based on the  conversion of adhesive 
filament interaction 
energy into force and which operates if individual filaments within 
a bundle are fully buckled in front of an obstacle as shown in 
fig.~\ref{figsnapshot}. 
The force generated by this mechanism is independent of 
chemical energy and entropic forces and is {\em not} limited by 
buckling. 
We characterize this zipping mechanism 
quantitatively and also show its intimate relation to a
force-induced unbinding transition of filament bundles.

\section{Bundle model}

In order to model a single bundle of $N$ filaments 
we start from an effective Hamiltonian containing bending energies
and interaction energies of all filaments,
\begin{equation}
{\cal H} = \sum_{i=1}^N{\cal H}_{b,i} + \sum_{i,j=1}^N {\cal H}_{2,ij}.
\label{H}
\end{equation}
In the first term,
 ${\cal H}_{b,i} = \int^{L_i}_{0} ds \frac{1}{2}\kappa (\partial_{s} 
{\bf t}_i)^{2}$  
is the bending energy of  filament $i$ with bending rigidity 
$\kappa$ and contour length $L_i$, 
which is parametrized by its arclength $s$ with a contour 
${\bf r}_i(s)$ and unit tangent vectors 
${\bf t}(s) \equiv \partial_s{\bf r}_i$.
We consider filaments with identical $\kappa$ and, thus, identical 
 persistence lengths $L_p \equiv \kappa/k_BT$ at temperature $T$.
The second term  describes  attractive
pairwise interactions between the filaments, 
$  {\cal H}_{2,ij} = \int_{0}^{\min(L_i,L_j)} ds
 [V_{\mathrm{r}}(\mathbf{\Delta r}_{\mathrm{ij}}) +
 V_{\mathrm{a}}(\mathbf{\Delta r}_{\mathrm{ij}})] $,
where $ \mathbf{\Delta r}_{\mathrm{ij}} = 
  \mathbf{r}_{\mathrm{i}}(s) - \mathbf{r}_{\mathrm{j}}(s)$
is  the distance between filaments $i$ and $j$ at arclength  $s$
along the filament. We assume that only monomers with 
similar arclength parameters interact. 
 The first
term is the hard-core repulsion of filaments with a potential 
$V_{\mathrm{r}}(\mathbf{r}) = \infty$ for $|\mathbf{r}| < \ell_{\mathrm{r}}$
and $V_{\mathrm{r}}(\mathbf{r}) = 0$ otherwise, where $\ell_{\mathrm{r}}$
is of order of the filament diameter. 
 The second term is an  short-range attractive
potential $V_{\mathrm{a}}(\mathbf{r})$, 
which we model by a potential well: the filament can gain an additional
energy $|W|$ over  a range $\ell_{\mathrm{a}}$,
\begin{equation}
V_{\mathrm{a}}(\mathbf{r}) = \left\{ 
   \begin{array}{l@{, \quad}l} 
        -|W| & \ell_{\mathrm{r}} \le |\mathbf{r}| <  
            \ell_{\mathrm{r}}+\ell_{\mathrm{a}}, \\ 
       0 & \mbox{otherwise}. 
    \end{array} 
  \right.
\label{potentiala}
\end{equation}
For cytoskeletal filaments,
 the attractive potential $V_a$ 
typically arises from   linker-mediated
attractions. Then  its strength $|W|$ is proportional to the 
crosslinker concentration 
in solution and the potential
 range $\ell_{\mathrm{a}}$ is  of the order of a
linker size. 
In the absence of forces and filament polymerization,
bundles of $N$ filaments form in a single discontinuous bundling transition 
at a critical potential strength $W_c^{(N)}$ \cite{kl03,kkl05}.

In the following we will consider bundles of polymerizing filaments 
which exert forces onto a planar wall.  We apply 
  clamped and capped boundary 
conditions on one end of the bundle ($s=0$), where all filaments 
are oriented into the $x$-direction and cannot polymerize or depolymerize. 
Because of the filament bending rigidity 
this induces a preferred orientation of the bundle into the $x$-direction,
see fig.~\ref{fig1}. 
Initially,  the capped ends $s=0$ of filaments are positioned in proximity.
The contour lengths  $L_i$ of filaments can differ because of the 
polymerization process. 
Monomers of length $\Delta l$ 
can attach and detach to and from  filaments at the other 
end ($s=L$) giving rise to changes in the contour length 
$\Delta L = \pm \Delta l$. The
attachment of monomers leads to a  polymerization 
energy gain $E_p<0$, which is related to  the ratio
$\omega_{\rm on}/\omega_{\rm off} = e^{-E_p/k_BT}$ 
of monomer attachment and detachment
rates. 
The attachment rate $\omega_{\rm on}$ is proportional 
to the monomer concentration in the surrounding solution and, thus, 
 also the  polymerization energy $E_p$ is controlled by  monomer 
concentration. 
We assume a constant  monomer concentration
 throughout the polymerization process, 
which implies a constant on-rate
$\omega_{\rm on}$ and, thus, constant $E_p$.

We will consider growth against a rigid planar wall
in the $yz$-plane perpendicular to the average filament orientation. 
The wall can move in the $x$-direction 
but cannot  rotate.
The wall is loaded with an additional force $F$.
We are not addressing ratchet mechanisms 
involved in the insertion of monomers at the loaded end of the 
filament in the proximity of the wall \cite{peskin93,mogilner96}.  
Therefore, we 
 assume that the wall has a very small diffusion coefficient
such that it moves
instantaneously with the position of the monomer 
with the maximal $x$-coordinate $x_{\rm max}$. 
Changes $\Delta x_{\rm max}$ give rise to an additional 
energy $F\Delta x_{\rm max}$  for the filament. 
If a monomer is attached such that 
$\Delta x_{\rm max}>0$, this leads to a
change in   the attachment rate 
$\omega_{\rm on}= e^{-(E_p+F \Delta x_{\rm max})/k_BT}\omega_{\rm off}$, where 
we assume that the value of $\omega_{\rm off}$ is unaffected by force.
Then 
growth  stalls for $F=|E_p|/\Delta x_{\rm max}$. 
The specific values of 
 $\omega_{\rm on,off}$  are not essential,  we only assume that 
shape fluctuations of the filaments are faster 
than their growth dynamics. In addition, 
 shape fluctuations can give rise  
to changes in $\Delta x_{\rm max}$ and corresponding energy changes.

\section{Buckling of single growing filaments}

A single cytoskeletal filament can generate forces
in the piconewton range \cite{peskin93,mogilner96}.
The polymerization force is defined  by  the corresponding   
load force  that stalls polymerization.
For a single filament this polymerization force is directly 
related to the polymerization energy $E_p$ per monomer, 
 $F_p = |E_p|/\Delta l$. 
 Polymerizing filaments will buckle if the load force 
$F$ exceeds the critical force for buckling,  $F_b \sim \kappa/L^2$. 
In the following we will discuss the possible dynamically 
stable steady states of 
growing or shrinking filaments.

For small load forces 
 $F<F_p$, the filament will grow, and the critical force for buckling,
$F_b \sim \kappa/L^2$, decreases. 
Eventually, the  load force $F$ becomes larger than  $F_b$,
and the filament buckles \cite{dogterom97}.
After  buckling, the growing filament 
end at $s=L$ has an angle $\phi_L>0$ with the $x$-axis, and
the polymerization force is opposed by 
the reduced load force $F\cos\phi_L$ in the direction tangential 
to the filament. 
This will lead to further growth, and the only stable state of a growing 
filament is the fully buckled state with  $\phi_L=\pi/2$.
Upon increasing the load force such that 
$F\cos\phi_L>F_p$, the filament shrinks. 
We find that the buckled  
state of a shrinking filament with $F_p = F\cos\phi_L<F$
represent an  {\em  unstable} mechanical equilibrium 
because $\phi_L$ is increasing for increasing $L$ at fixed load force $F$. 
For flexible walls, a similar instability has been discussed in 
  Ref.\ \cite{daniels06}.
Therefore,  the only stable states of a 
 shrinking filament are the unbuckled state 
 ($\phi_L=0$)  and the fully buckled state ($\phi_L=\pi/2$)
 as long as the length reservoir is 
sufficiently large. 
Because growing filaments will 
always end up in a fully buckled state, 
 mechanisms for  force generation which 
also operate in the fully buckled state of individual filaments
are essential in systems containing polymerizing filaments.
We will demonstrate
 that filament bundles can generate forces  
using a zipping mechanism
if  each filament in the bundle is
 fully buckled.

\section{Zipping and force-induced unbinding}

Cells usually rely on bundles of several filaments for 
the formation of cell protrusions such as 
filopodia or acrosomal extensions. Such bundles 
have a higher bending rigidity  \cite{bathe08} 
and are more stable against buckling.
Stall forces of  polymerizing actin bundles could be determined
experimentally only recently \cite{footer07}. 
If the force generation mechanism is based on the polymerization energy 
$E_p$ of single filaments, bundles of filaments 
 are believed to have higher stall forces because of 
load sharing. 
It has been proposed that within
bundles filaments can additionally exploit an attractive interaction 
 to generate higher forces \cite{mahadevan00,MO03}. 

Within this Letter, we quantitatively investigate the 
interplay of attractive 
bundling interactions and  external load force. 
We find that it is possible
to generate forces independently of the polymerization energy $E_p$ and 
entirely based on the  attractive interaction  between filaments 
 by a zipping mechanism. 
In this mechanism, the adhesive energy which is  
gained during bundle formation generates a zipping force.

\begin{figure}
  \begin{center}
  \epsfig{file=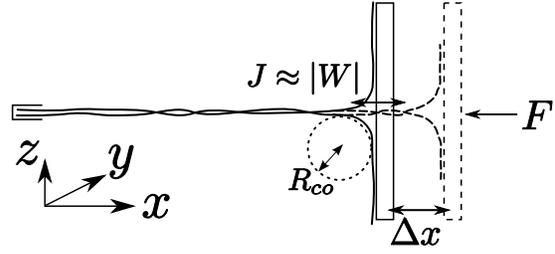,width=0.4\textwidth}
  \caption{
   Zipping mechanism in front of a wall with load force $F$. 
   Zipping starts in a splayed configuration of filament ends 
   (solid lines). The curvature at the wall is given by the contact radius 
  $R_{co}$. 
   Zipping a distance $\Delta x$ (dashed lines) 
      performs a work  $F\Delta x$  but gains 
  an adhesive energy $J\Delta x$.
   }
  \label{fig1}
  \end{center}
\end{figure}

We will first explain the mechanism for two filaments.
As shown in fig.~\ref{fig1},
zipping of two filaments requires a particular initial 
condition, a ``zipping fork'', 
where both filaments are in a fully buckled state 
with   $\phi_L= \pi/2$ and  well-separated  uncapped 
 filament ends at the wall in  a splayed 
configuration.
As explained above, the fully buckled state is generic for
 polymerizing non-interacting filaments.
 The splayed initial condition arises then naturally
by the thermal  motion of uncapped filaments  ends if the capped ends 
are anchored in proximity and
the crosslinker concentration or the adhesive potential 
is increased from low values ($|W|< |W_c^{(2)}|$).
The wall exerts a total force $F$ in
the negative $x$-direction. If the two filaments bind 
together along an
additional length $\Delta x$, the bundle gains the
  free energy $J\Delta x$, 
where $J > 0$ represents the free energy of bundling, which 
arises from the competition of thermal shape  fluctuations of filaments 
and the short-range attraction between filaments \cite{kkl05}. 
This implies that the zipping mechanism will only work in the 
bundled phase. 
In the absence of thermal fluctuations,
we have  $J = |W|$, i.e.\ the bundling free energy $J$ equals the 
potential interaction  energy gain $|W|$. 
In the presence of thermal shape fluctuations, the potential
energy is reduced by entropic contributions. 
Close to the discontinuous unbinding transition, the
free energy vanishes according to 
 $J \sim |W_c^{(2)} - W|$. \cite{kkl05} 
If the filaments bind
together along an additional length $\Delta x$, the wall has to 
move the same distance  $\Delta x$ against the load 
force $F$. This movement performs a work $F \Delta x$, and the total 
free energy gain is 
\begin{equation}
  \Delta G = (J - F) \Delta x,
\label{deltaG}
\end{equation}
see fig.~\ref{fig1}.
 If the load force $F$ is
smaller than the  critical force $F_c^{(2)} =J$, 
a change $\Delta x>0$ of the bound length leads to 
  a free energy gain $\Delta G >0$ 
resulting in  spontaneous  {\em zipping}. 
The critical force $F_c^{(2)}$ represents the maximal
force which can be generated by the zipping mechanism for two filaments. 
For forces $F > F_c^{(2)}=J$, 
an ``inverse'' zipping with $\Delta x<0$ leads to a 
free energy gain, i.e.\ the bundle is 
separated by the load force $F$. 
This process represents a  {\em force-induced unbinding}. 
 Deep inside the bundled phase, i.e.\
for  $|W| \gg |W_c^{(2)}|$, we find critical forces 
  $F_c^{(2)} = J \approx |W|$. Close to the
thermal unbinding transition the critical zipping force vanishes as 
$F_c^{(2)} = J \sim |W_c^{(2)} - W|$. 

We can  also consider zipping and 
force-induced unbinding as a function of the potential strength $W$ for 
fixed force $F$. Force-induced unbinding then happens for $|W|$ smaller than 
 a force-dependent critical 
potential strength $W_c^{(2)}(F)$, which 
is given by  $|W_c^{(2)}(F)|\approx F$ deep in the bundled 
phase, where large forces are needed to unbind the bundle, 
 and approaches the 
critical potential strength for purely thermal unbinding,
$W_c^{(2)}(F)\approx  W_c^{(2)}-F$,  for small forces. 
Zipping takes place above the critical potential strength for 
 $|W|>W_c^{(2)}(F)$.

All zipping and force-induced unbinding thresholds are independent 
of the polymerization energy $E_p$ and, thus,  these phenomena 
do not depend on the 
presence of the polymerization force. 
The zipping mechanism  exploits  the binding free energy $J$ between 
filaments. The polymerization at the end of
the filaments is needed only  to provide  a sufficient
reservoir of length for the bundle such that force  
can be generated continuously. 
The polymerization has to be sufficiently fast to establish a 
length reservoir but the 
details of the polymerization kinetics are not important for 
the zipping mechanism.

The mechanism  requires the separation of filament ends 
in  the splayed zipping fork configuration
in order  to  avoid binding of filaments 
by  rotation around the $x$-axis without any  force generation.
This separation is maintained by the 
 slow kinetics of the long 
polymer ends or by fixing the filament ends in the $yz$-plane.
In the splayed configuration semiflexible filaments attain 
a  radius of curvature at the wall, which is given by the 
contact radius $R_{co}\sim (\kappa/J)^{1/2}$ \cite{K06}, see fig.~\ref{fig1}. 
The stiffness of filaments is  important in order to allow for a force
transmission onto the wall by the curved contact segments. 
Only filaments with a nonzero  bending rigidity 
  can exert a torque onto the wall  in the fully buckled state.

\section{Monte Carlo simulations}

In order to gain further insight into  zipping and force-induced unbinding 
for $N\ge 2$ filaments we 
 have performed  
Monte Carlo (MC) simulations for
identical filaments using the  effective Hamiltonian (\ref{H}).
Simulation snapshots are shown in fig.~\ref{figsnapshot}.
In the MC simulation we use a discretized 
parameterization in terms of the arc length 
$s$ as in the worm-like chain model 
 ${\cal H}_{b,i} = \int^{L_i}_{0} 
ds \frac{1}{2}\kappa (\partial_{s} {\bf t})^{2}$ 
and model the constraint $|{\bf t}(s)|=1$  by a sufficiently 
stiff harmonic potential. 
The contours ${\bf r}_i(s)$ of each filament $i$ of 
length $L_i$ are discretized into 
$M_i = L_i/\Delta s$ equidistant points
 $\mathbf{r}_{i}^n={\bf r}_i(n\Delta s)$. 
The total energy 
${\cal H} = \sum_{i}{\cal H}_{b,i} + \sum_{i,j} {\cal H}_{2,ij}$
 is calculated using  a
discretized bending energy 
$ {\cal H}_{b,i} =  \sum_{n=1}^{M_i} \kappa \left( 1 - 
    \hat{\mathbf{r}}_{i}^{n-1,n} \cdot 
    \hat{\mathbf{r}}_i^{n,n+1} \right) 
       + k (|\mathbf{r}_i^{n,n+1}| - \Delta s)^2$,
where $\mathbf{r}_i^{n-1,n} \equiv\mathbf{r}_i^n - \mathbf{r}_i^{n-1}$ and 
$\hat{\mathbf{r}}_i^{n-1,n} \equiv 
\mathbf{r}_i^{n-1,n}/|\mathbf{r}_i^{n-1,n}|$,
and the second term represents the spring
energy that enforces the constraint 
$|{\bf t}_i(s)|=1$ in the discretized model (we use $k=100k_BT/\Delta s^2$). 
We also discretize the 
attractive interaction energy according to 
$ {\cal H}_{2,ij} =\sum_{n=1}^{\min{M_i,M_j}} 
[V_{\mathrm{r}} (\mathbf{r}_{ij}^n) + V_{\mathrm{a}} (\mathbf{r}_{ij}^n)]$.
The effects of monomer attachment and detachment and the load force 
can be taken into account by additional energy contributions
${\cal H}_p = E_p \sum_i M_i$ and ${\cal H}_F = F x_{\rm max}$.

We employ the Metropolis algorithm for the 
 total energy ${\cal H}+{\cal H}_p+{\cal H}_F$.
For  configurational equilibration
 we offer local displacement moves of the vectors ${\bf
  r}_i^n$ in each MC step and  pivot moves of whole filament segments. 
In addition we attempt attachment and detachment moves of monomers with 
smaller frequency in order to achieve  
shape fluctuations of filaments which are faster 
than the growth dynamics. 
For  a fast equilibration
for longitudinal fluctuations of the bound or zipped length 
along the filament  we also attempt reptation-like moves where monomers 
are transferred between the  capped end at $s=0$ and the uncapped end 
at $s=L_i$ and vice versa; these moves do not change 
the total number of monomers. 
The zipping mechanism  relies on the separation of filament ends 
into a split zipping fork configuration
at the wall, see fig.~\ref{figsnapshot}. Filament ends 
  have to stay separated in order to
 avoid binding of filaments  by  simple rotation. 
In the MC simulations such rotations are kinetically suppressed 
as the rotational diffusion of a whole filament by local 
displacement moves happens on 
much larger time scales as  zipping, which is accelerated in the MC 
simulations by the reptation-like moves.

\begin{figure*}
  \begin{center}
  \epsfig{file=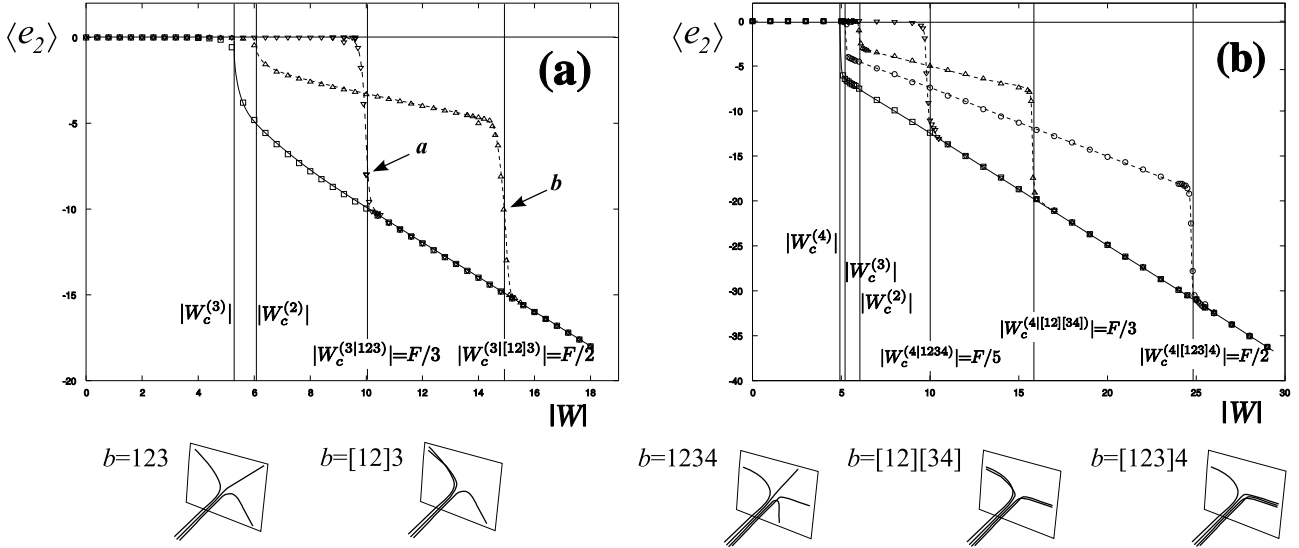,width=0.95\textwidth}
  \caption{
     MC data for the average binding energy 
     per filament and per length, $\langle e_2 \rangle$ 
    as a function of the potential strength $|W|$ for 
    (a) $N=3$ and (b) $N=4$ identical filaments (with 
     persistence length $L_p = 100$, initial contour length $L=100$, 
     potential range  
    $\ell_{\mathrm{a}}=0.001$ and hard-core radius $\ell_{\mathrm{r}}=0.1$; 
     all lengths in units of $\Delta l$; energies in units of $k_BT$;
   lines are guides to the eye).
     Arrows correspond to the snapshots in fig.~\ref{figsnapshot}.
    In the absence of an
    external force $F = 0$ ($\Box$), the thermal unbinding 
      transition  happens at a critical potential strength 
    $|W_{c}^{(N)}|$. 
     (a) 
    For $N=3$ an external force $F = 30$ is applied.
       For  an initial configuration $b=123$ 
    ($\bigtriangledown$),
    the unbinding transition occurs at a  critical potential strength 
     $|W_{c}^{(3|123)}| \approx F/3$. 
    For  an initial condition $b =[12]3$ ($\bigtriangleup$), 
    a cascade of two  unbinding transition occurs
    at  critical potential strengths $|W_{c}^{(2)}|$ and 
    $|W_{c}^{(3|[12]3)}|\approx F/2$. 
    (b) 
       For $N=4$ filaments an 
     external force $F=50$ is applied.
    This leads to three different force-dependent critical 
     potential strengths $|W_c^{(4|1234)}|\approx F/5$ ($\bigtriangledown$), 
    $|W_c^{(4|[123]4)}|\approx F/3$ ($\circ$), and 
    $|W_c^{(4|[12][34])}|\approx F/2$ ($\bigtriangleup$) depending on the 
    initial subbundle configuration.
   }
  \label{fig2}
  \end{center}
\end{figure*}

\section{Force-induced unbinding transition}

We first  consider  the force-induced
unbinding transition of filament bundles.
In fig.~\ref{fig2}, we show MC results for 
the average binding energy  per 
 length and per filament, 
 $\langle e_2 \rangle \equiv
\langle (\sum_{i,j=1}^N{\cal H}_{2,ij})/(\sum_{i=1}^N L_i) \rangle$,
for bundles with $N=3$ and $N=4$ 
 in the presence of a load 
force $F$ and as a function of the 
potential strength per length $|W|$.

In the absence of external forces,
a single, discontinuous 
unbinding transition occurs at a critical potential strength $W_c^{(N)}$,
which only depends on
the number of filaments in the bundle ~\cite{kkl05}. In the presence of
a load  force, on the other hand, 
the unbinding transition occurs (i) in several steps, (ii) 
at critical potential strengths, which depend on the load force,
and (iii) via different pathways depending on the 
 initial subbundle configuration. 
The number of transition steps and the critical potential strengths in 
 force-induced unbinding depend 
on the initial zipping fork configuration, 
in particular on the number and types  of 
subbundles in the initial splayed configuration.
For $N>2$ filaments several initial subbundle configurations are possible. 
We will focus on  conditions deep in the bundled phase of  $N$ 
filaments. Then, starting with high potential strengths $|W|$, 
we first 
 find  a force-induced unbinding of subbundles 
at a critical potential strength $|W_c^{(N|b)}(F)|\approx F/n(N|b)$,
 where $b$ will 
 index  the initial subbundle configuration and with 
a number $n(N|b)$ of pairwise filaments interactions 
lost upon subbundle unbinding. 
Then, at  smaller 
critical potential strengths $|W_c^{(M)}|$,
there is  a subsequent 
thermal unbinding transition of subbundles containing $M$ filaments, 
which is independent of force.

For a bundle with $N=3$ filaments, two different initial zipping fork
configurations and, thus, two force-induced unbinding pathways 
 are possible, see fig.~\ref{fig2}(a). In configuration $b=123$,
all three filaments 
 point in different directions. In configuration $b=[12]3$, the end
of the bundle is split into one subbundle of two bound filaments $[12]$ 
and the  third filament $3$ pointing  in a different direction. 
In configuration $b=123$, there is a single unbinding transition 
at $|W_c^{(3|123)}| \approx  F/3$ with  
$n(3|123)=3$ pairwise filament interactions 
lost upon unbinding. 
In configuration $b=[12]3$, on the other
hand, there are two unbinding transitions: First, filament 3 is separated 
from the subbundle $[12]$ at $|W_c^{(3|123)}| \approx  F/2$ because  
$n(3|123)=2$  pairwise filament interactions are 
lost upon subbundle unbinding. Further decreasing the potential 
strength $|W|$, there is a second thermal unbinding transition 
of the subbundle $[12]$ at the critical value $|W_c^{(2)}|$, which 
is independent of force. The values for $n(3|b)$ 
correspond to a  triangular arrangement of a three filament 
bundle, as can be seen in fig.~\ref{figsnapshot}. 

For bundles with $N>3$ even more initial subbundle configurations 
are possible giving rise to a variety of possible force-induced 
unbinding pathways. In fig.~\ref{fig2}(b), we show MC results for 
a bundle with $N=4$ filaments, which exhibits already 
three different unbinding pathways under force. 
These pathways are related to the 
initial configurations  $b=1234$ with  four 
separated filaments, 
$b=[12][34]$ with two subbundles containing two filaments each, and 
$b=[123]4$ with one subbundle containing three filaments and one separated 
filament. 
The numbers of pairwise filament interactions 
lost upon unbinding are $n(4|1234)=5$, $n(4|[12][34])=3$, and 
$n(4|[123]4)=2$. All three values for $n(4|b)$ can be  explained by a 
triangular arrangement of filaments in the bundle, as it has been  
observed for equilibrium bundles in Ref.\ \cite{kkl05}.
After force-induced-unbinding, the remaining 
subbundles unbind thermally at  lower
critical potential strengths in a second transition.

\section{Zipping}

Whereas a bundle of $N$ filaments unbinds for $|W|<W_c^{(N|b)}(F)$, 
it starts  zipping 
above  the critical potential strength, for   $|W|>W_c^{(N|b)}(F)$.
The filament  can  generate and transmit 
forces onto  a wall by the zipping mechanism, which  converts
adhesive energy into a force. 
The critical  force $F_{c}^{(N|b)}$ is the maximal force that  a bundle 
with $N$ filaments and initial conditions $k$ can generate 
by the zipping mechanism  for  
a given potential strength $|W|$.
The  critical forces $F_c^{(N|b)}$ for zipping with an initial 
condition $b$ are related  to the critical potential strengths
$W_c^{(N|b)}(F)$ for force-induced unbinding by 
$ W_c^{(N|b)}(F_c^{(N|b)})= |W|$, which gives 
$F_c^{(N|b)} \approx |W|/n(N|b)$   at high potential strengths.
Therefore, we also find  different 
critical zipping forces depending on the zipping pathway, which is determined 
by the initial configuration $b$.

The kinetics of zipping in the MC simulation is characterized by the 
 average velocity $\langle v_w \rangle$ of the wall 
along the $x$-axis in the stationary state. 
For a given load force $F$, the velocity $\langle
v_w \rangle $ changes sign at the critical potential 
strength   $W_c^{(2)}(F)\approx F$ with 
 $ \langle v_w \rangle < 0$ for force-induced unbinding
for  $|W|<W_c^{(2)}(F)$  and  $ \langle v_w \rangle > 0$ 
for zipping for $|W|>W_c^{(2)}(F)$. 
The average velocity is given by 
$ \langle v_w \rangle = (\omega_+ - \omega_{-}) \Delta l $
in terms of 
 the  rates $\omega_{+}$ and $\omega_{-}$  for zipping and 
unzipping a segment of length $\Delta l$. 
 These  rates depend 
on the attempted MC moves,  and  their sum $\omega_0=\omega_+ +\omega_-$ 
 is  given  by 
the frequency at which  reptation-like moves are 
offered in the  MC dynamics.
Furthermore, eq.\ \ref{deltaG}  leads to  $\omega_+/\omega_{-}= 
\exp{[(J-F) \Delta l/k_BT]}$ 
with $J-F \approx |W| - |W_2^{(2)}(F)|$,
such that
\begin{eqnarray}
\langle v_{w} \rangle 
   &\approx& v_0 
    \tanh\left(\frac{(|W| - |W_2^{(2)}(F)|) \Delta l}{2k_BT} \right)
\label{vw}
\end{eqnarray}
with a maximal velocity $v_0 = \omega_0 \Delta l$.
This result 
is in agreement with results from our MC simulations
(data not shown) and shows that  
the  width of the transition between 
force-induced unbinding an zipping decreases with 
decreasing temperature $T$.

In the MC kinetics we neglect frictional forces, which
limit reptation-like motion.
In a real system we expect  the result (\ref{vw}) for the 
velocity-potential relation  to hold for 
$|\langle v_w \rangle|\ll  v_0$, i.e.,
 close to equilibrium with 
a maximal velocity $v_0$, which is determined by the equilibrium 
between zipping force and frictional force  of the polymer ends.

\section{Discussion and Conclusion}

We have shown that  forces can be generated by a  zipping mechanism,
which is completely based on the  conversion of adhesive 
filament interaction 
energy into force and which operates independently of the polymerization 
energy if filaments within 
a bundle are fully buckled. 
Below a  critical potential strength or above a critical 
load force  zipping does no longer occur, and 
there is a transition  from zipping to a 
 force-induced unbinding of the filament bundle.

The resulting zipping force is given by the filament 
interaction energy per length which is liberated upon separating 
the filaments. For F-actin crosslinkers such as $\alpha$-actinin or filamin 
recent measurements give binding energies of $4k_BT$ per 
crosslinker and filament pair \cite{ferrer08}, which yields 
zipping forces  $F_c^{(2)} \simeq 6 {\rm pN}$ 
 for two filaments if we assume one crosslinker per actin monomer. 
Actin filaments can also be bundled by counterions with 
typical binding energies of  $0.02k_BT$ per actin monomer for magnesium 
ions \cite{tang01}, which are much weaker than  protein crosslinkers. 
 For these interactions bundles of the 
order of $N=10$ filaments are needed to generated zipping forces 
in the piconewton range if we assume a triangular filament arrangement 
and separation into single filaments resulting in $n^{(N|b)} \approx 3N$
for large $N$.

The zipping mechanism only relies on adhesive energy and 
does not require a large variety of regulatory proteins as
found for  actin-based motility of eukaryotic cells 
 \cite{pollard03,carlier03}.
Zipping  mechanisms may also contribute to force generation
in the presence of regulatory proteins, in particular, force generation 
by filament bundles in cell protrusions such as filopodia \cite{MR05}
but they
 could  play a more prominent important role for 
the motility of relatively primitive cells such as  sperm cells of nematodes
~\cite{MO03,miao03,roberts00}.
Zipping  mechanisms could also be exploited to create 
artificial force generating systems 
using synthetic semiflexible polymers with attractive interactions.

\acknowledgments
We  acknowledge financial support within
the Collaborative Research Center ‘‘Mesoscopically
Structured Composites’’ (SFB 448) of the Deutsche
Forschungsgemeinschaft.


\end{document}